\documentclass[twocolumn,pra,superscriptaddress,showpacs,aps,floatfix]{revtex4-1}
\usepackage{color}
\usepackage{amsmath}
\usepackage{amssymb}
\usepackage{txfonts}
\usepackage{graphicx}
\usepackage{epstopdf}
\usepackage[unicode]{hyperref}
\hypersetup{pdftitle={2D SOC BEC},
 pdfauthor={},
 pdfsubject={2D SOC BEC},
  colorlinks=true,
    linkcolor=red,
    citecolor=magenta,
    filecolor=black,
    urlcolor=blue}


\begin{document}

\title{A two-Higgs-doublet Model without flavor-changing neutral currents at tree-level }

\author{Chilong Lin}
\email{lingo@mail.nmns.edu.tw}

\affiliation{National Museum of Natural Science, 1st, Guan Chien RD., Taichung, 40453 Taiwan, ROC}

\date{Version of \today. }

\begin{abstract}

The flavor-changing neutral current (FCNC) problem at tree-level is a very critical defect of the two Higgs doublet extension of standard model (SM).
In this article, a two-Higgs-doublet model (2HDM) in which such defects do not exist at all is to be demonstrated.
The general pattern of matrix pairs which can be diagonalized simultaneously by a same unitary transformation is proposed without extra constraints like symmetries or zeros in $M$ matrices.
Only an assumption of the hermiticity of mass matrices is employed in the derivation.
With this assumption, number of parameters in the mass matrix of a specific fermion type is reduced from eighteen down to five.
Eigenvalues and eigenvectors are analytically derived and it¡¦s surprising that unitary transformation matrix thus derived depends on only two parameters.
It is a very general and elegant way to solve the tree-level FCNC problem radically and it includes previous similar models as special cases with specific parameter values.

\end{abstract}
\maketitle


\section{Introduction}

In the standard model (SM) of electro-weak interactions CP symmetry can be violated explicitly by ranking the Yukawa couplings between fermions and Higgs fields suitably and expect them to generate complex phases in the Cabibbo-Kobayashi-Maskawa (CKM) matrix.
But, no one knows how these Yukawa couplings should be arranged to give a satisfactory CKM matrix.
 Though in SM the vacuum expectation value (VEV) of its Higgs doublet also provided a complex phase when the gauge symmetry was spontaneously broken. 
However, CP symmetry can't be violated spontaneously in SM since the only phase generated in this way can always be rotated away. Thus, an extension of SM with one extra Higgs doublet was proposed \cite{TDLee1973}.
Through which, one expects the phases in the VEVs may unlikely be rotated away simultaneously and a non-zero phase difference between them might survive so as to bring in CP-violation spontaneously.
However, a definite way regarding how CP-violation can be generated in such a model is still missing. \\

Besides failing to solve the CP problem, this extra Higgs doublet also brought in an extra problem of flavor-changing neutral currents (FCNCs) at tree level.
This problem arises if those two components $M_1$ and $M_2$ of the quark mass matrix  $M=M_1+M_2$ corresponding to Higgs doublets $\Phi_1$ and$\Phi_2$ respectively were not diagonalized simultaneously by a same unitary transformation $U$.
Respective non-zero off-diagonal elements of diagonalized $U M_1 U^{\dagger}$ and $U M_2 U^{\dagger}$ in the mass eigenstate lead to flavor-changing-neutral (FCN) interactions mediated by neutral Higgs scalars at tree level. \\

At the beginning, people were not aware the danger of such interactions too much. But, as the energy and accuracy of experiments increase and no such effects were detected until 2005 \cite{Acosta2005}, the need of hypotheses to explain the smallness of such interactions emerged.
Owing to the smallness of detected such interactions, it is natural for people to consider them as loop corrections.
If one considers them as loop corrections, that means such interactions won't appear at tree-level.
In such a manner, there should be matrix pairs which can be diagonalized simultaneously and how to find such matrix pairs becomes the key to solve the FCNC problem. \\

In \cite{Glashow1977} two Natural-Flavor-Conservation (NFC) models were proposed by employing a $Z_2$ discrete symmetry to forbid both Higgs doublets to couple with a same quark type simultaneously.
These two models are usually referred to as the Type-I and Type-II NFC models or just Type-I and Type-II 2HDMs.
In the Type-I model, only one of the Higgs doublets couples with both quark types and the other completely does not.
In the Type-II model, up- and down-type quarks couple with different Higgs doublets respectively.
Surely they are free of FCNCs since for a specific quark type either $M_1 =0$ or $M_2 =0$ guaranties the absence of FCNCs.
Besides these two models, there are also similar models which are usually referred to as the Type-III and -IV models,
sometimes the -X and -Y models, or the lepton-specific and flipped 2HDMs.
However, these models are just extensions of Type-I and -II models to include leptons~\cite[and references therein]{Aoki2009, Pich2009, Branco2012, Diaz-Cruz2004}. \\

Besides models mentioned in last paragraph, there are also several 2HDMs in which both $M_1$ and $M_2$ are assumed non-zero.
But these models do not have proper dynamic explanations for their assumptions.
For instance, in Fritzsch-ansatz ~ \cite{Fritzsch1978, Fritzsch1979} and its consequent developments several elements of the mass matrices were assumed zero to simplify the pattern of mass matrices down to an analytically manageable level.
In the Aligned Two-Higgs-Doublet model (ATHDM or A2HDM) \cite{Pich2009}, an assumption of $M_1$ and $M_2$ are proportional was employed as an $ad-hoc$ constraint.
However, they are in fact special cases of the model to be presented in this article.
It is noticeable that most of these models also assume the $M$ matrices are Hermitian in addition to these imposed constraints. \\

Besides the FCNC-free 2HDMs mentioned above, there is another type of 2HDM which is also referred to as Type-III 2HDM~\cite[and references therein]{Cheng1987, Liu1987, Atwood1997, Buras2010, Sanchez2013, Crivellin2013} and usually causes confusions with the one mentioned above.
Such models do not deny the existence of FCNCs at tree level.
It just assumes that tree-level FCNCs are highly suppressed or can be canceled by loop corrections down to empirical values. \\

In 2HDMs mentioned above, except for the one allows for tree-level FCNCs, assumptions or forbiddance were employed to simplify the patterns of fermion mass matrices.
However, theoretically, a model with fewer $ad-hoc$ constraints is better since it will be more general and natural.
Forbidding couplings between specific fermion types and Higgs doublets is a very strong constraint.
In usual, physicists prefer a general model applies to wider fields to special models constrained by special conditions or symmetries.
Thus, a model in which neither $M_1$, $M_2$ nor any elements in them has to be assumed zero will be a better one. \\

Theoretically, a 2HDM in which both $M_1$ and $M_2$ are nonzero and can be diagonalized simultaneously should exist.
If such a model does exist, it is obviously more general than the $Z_2$-symmetric ones or those assuming zeros in mass matrices.
At the mean time, assumptions of highly suppressed tree-level FCNCs or loop corrections as strong as zeroth-order interactions in Type-III models are no longer needed.
In such a model, FCNCs at tree-level are no more problems since they do not exist at all.
Now the only problem is how to find them and what will they looks like. \\

In fact, such matrix pairs do exist and the first one had been discovered in a $S_3$-symmetric 2HDM ~\cite{Lin1988, Lee1990, Lin1994} decades ago.
The matrix pair derived there were
\begin{eqnarray}
M_1 = \left( \begin{array}{ccc} A & B & B   \\  B & A & B  \\ B & B & A \end{array}\right),
~~~~M_2 = i \left( \begin{array}{ccc} 0 & -D & D   \\  D & 0  & -D  \\ -D & D & 0 \end{array}\right),
\end{eqnarray}
and the unitary transformation which diagonalize them simultaneously was derived analytically as:
\begin{eqnarray}
U = \left( \begin{array}{ccc}
1/\sqrt{3} & (-1-i \sqrt{3}) /2\sqrt{3} & (-1+i \sqrt{3}) /2\sqrt{3} \\
1/\sqrt{3} & (-1+i \sqrt{3}) /2\sqrt{3} & (-1-i \sqrt{3}) /2\sqrt{3} \\
1/\sqrt{3} & 1/\sqrt{3} & 1/\sqrt{3}
\end{array}\right).
\end{eqnarray}

In recent years, three more such FCNC-free matrix pairs were derived by extending the $S_3$-symmetric pattern to three $(S_3 +S_2)$-symmetric patterns ~\cite{Lin2013}.
In this article, a very general pattern of such FCNC-free matrix pairs is derived without any symmetries or assumptions except the hermiticity of mass matrices.
It not only includes the $S_3$- and $(S_3+S_2)$-symmetric matrix pairs as special cases in it,
but also the $Z_2$-symmetric, Fritzsch ansatz and A2HDM. \\

At the beginning of section II, a FCNC-free condition for a 2HDM which was firstly given in \cite{Branco1985} is discussed and improved.
With this condition, number of free parameters in a mass matrix is substantially reduced from eighteen down to five.
Thus, analytical solutions of the eigenvalues and eigenvectors are derivable.
It is amazing that derived unitary transformation matrix $U$ which diagonalize both $M_1$ and $M_2$ simultaneously is extremely simple.
It depends on only two of the five parameters in each quark type.
Conclusions and some discussions on its application on generating CP violation will be given in section III. \\

\section{A General Pattern for FCNC-free Mass Matrices}

Theoretically, finding matrix pairs which can be diagonalized by a same $U$ matrix simultaneously is a better way to solve the FCNC problem since that will be more general than those imposing symmetries ~\cite{ Lin1988, Lee1990, Lin1994, Lin2013}, forbiddance ~\cite{Glashow1977} or zeros in mass matrices \cite{Fritzsch1978, Fritzsch1979}.
In usual, physicists prefer a general model applies to wider fields to special models apply only under certain conditions. \\

Following this hypothesis, we would like to start the study from a very general basis without any imposed symmetries, forbiddance or zeros.
We try to keep it as general as possible in the following derivations.
However, as to be shown below, a Hermitian assumption of the $M$ matrix is still needed for simplifying the $M$ matrix down to a manageable level.
This Hermitian assumption firstly reduce the number of parameters in a most general $M$ matrix from eighteen down to nine.
Then, with the help of an interesting condition between $M_1$ and $M_2$ which is to appear in Equation (7), the parameter number is further reduced down to five.
Thus, eigenvalues and corresponding $U$ matrix are now achievable analytically.
In what follows, the procedure of derivations will be presented step by step.  \\

For a 2HDM, the Yukawa couplings of $Q$ quarks can be written as
\begin{equation}
-{\cal L}_Y ~=  \bar{Q_L} (Y^d_1 \Phi_1 +  Y^d_2 \Phi_2) d_R +  \bar{Q_L} \epsilon (Y^u_1 \Phi_1^{\ast} + Y^u_2 \Phi_2^{\ast} ) u_R + h. c.,
 \end{equation}
where $Y^q_i$ are $3 \times 3$ Yukawa-coupling matrices for quark types $q=u$, $d$ and Higgs doublets $i=1$, $2$, respectively, and $\epsilon$ is the $2 \times 2$ antisymmetric tensor.
$Q_L$ are left-handed quark doublets, and $d_R$ and $u_R$ are right-handed down- and up-type quark singlets, respectively, in their weak eigenstates.
The mass matrices can then be expressed as
\begin{equation}
M^{(u,d)}=M^{(u,d)}_1+M^{(u,d)}_2=Y^{(u,d)}_1 \langle \Phi_1 \rangle + Y^{(u,d)}_2 \langle \Phi_2 \rangle ,
 \end{equation}
where $\langle \Phi_i \rangle \equiv  v_i e^{i \theta_i}/ \sqrt{2}$, $i=1$ and $2$, are vacuum expectation values of the Higgs doublets.
Here, a basis in which $\theta_1 =0$ will be chosen and only the phase $\theta_2$ is left non-zero. \\

Neglecting the hyper-indices of $M$, the most general pattern of a $3\times 3$ mass matrix can always be written as
\begin{eqnarray}
M &=& M_1 +M_2 = \left( \begin{array}{ccc} A_1 +i D_1 & B_1+ i C_1 & B_2+ i C_2   \\
 B_4 + i C_4 & A_2 +i D_2 & B_3 + i C_3  \\ B_5 + i C_5 & B_6 + i C_6 & A_3 +i D_3 \end{array}\right),
\end{eqnarray}
where $A$, $B$, $C$ and $D$ are all real and each of them may receive contributions from both $M_1$ and $M_2$ arbitrarily.
Such a pattern is the most general one for a $3 \times 3$ matrix since it contains nine elements and each of them has one real and one imaginary components.
Thus, if no constraints were imposed, there are eighteen parameters in such a $M$ matrix in total. \\

It is obvious that eighteen parameters are too many to have such a matrix be diagonalized analytically.
That's why physicists employed various constraints in previous 2HDMs to simplify the $M$ pattern.
However, as to be shown below, a Hermitian assumption is already enough to simplify the $M$ pattern down to an analytically manageable level.
Constraints like  $Z_2$, $S_N$ symmetries or imposed zeros in $M$ matrix are in fact unnecessary. \\

If one assumes the mass matrix $M$ were Hermitian, the $D$ parameters will be all zero and Equation (5) becomes
\begin{eqnarray}
M &=&  \left( \begin{array}{ccc} A_1  & B_1+ i C_1 & B_2+ i C_2   \\
 B_1 - i C_1 & A_2  & B_3 + i C_3  \\ B_2 - i C_2 & B_3 - i C_3 & A_3  \end{array}\right),
\end{eqnarray}
with $B_4=B_1$, $B_5=B_2$, $B_6=B_3$, $C_4=-C_1$, $C_5=-C_2$ and $C_6=-C_3$. \\

In ~\cite{Branco1985}, a condition
\begin{equation}
M_1 M_2^\dagger - M_2 M_1^\dagger = 0,
 \end{equation}
was given for a matrix pair which can be diagonalized by a same $U$ matrix if they were both Hermitian.
This can be easily proved since $U M_1 U^{\dagger} = U M_1^{\dagger} U^{\dagger} =M_1^{diag.}$ and $U M_2 U^{\dagger}= U M_2^{\dagger} U^{\dagger} =M_2^{diag.}$,
 where $M_1^{diag.}$ and $M_2^{diag.}$ are diagonal.
If one applies $U$ onto Equation (7), one will receive
\begin{eqnarray}
&( & U M_1 U^{\dagger}) ( U M_2^{\dagger} U^{\dagger}) - (U M_2 U^{\dagger})( U M_1^{\dagger} U^{\dagger}) \nonumber \\
 &=& M_1^{diag.} M_2^{diag.} -M_2^{diag.} M_1^{diag.} =0.
\end{eqnarray}
Thus, if one can find a matrix pair which satisfies Equation (7), the FCNC problem at tree-level is solved automatically. \\

Based on this, it is instinctive for one to divide the Hermitian matrix in Equation (6) into two Hermitian components and substitute them into Equation (7).
The simplest way for doing so is to divide them into one purely real component and one purely imaginary component as
\begin{eqnarray}
M_R = \left( \begin{array}{ccc} A_1 &  B_1 &  B_2  \\  B_1 &  A_2 &  B_3  \\ B_2 & B_3 & A_3 \end{array}\right) ,  ~~~~~
M_I = i \left( \begin{array}{ccc} 0 & C_1 & C_2   \\  -C_1 & 0 & C_3  \\ -C_2 & -C_3 & 0 \end{array}\right).
\end{eqnarray}

Substitute Equation (9) into Equation (7), we receive
\begin{eqnarray}
 M_I M_R^\dagger &=& i \left( \begin{array}{ccc} B_1 C_1 +B_2 C_2 & A_2 C_1+B_3 C_2 & B_3 C_1+A_3 C_2  \\ B_2 C_3-A_1 C_1 & B_3 C_3-B_1 C_1 &  A_3 C_3-B_2 C_1  \\ -A_1 C_2-B_1 C_3 & -B_1 C_2-A_2 C_3 &  -B_2 C_2-B_3 C_3 \end{array}\right),  \nonumber \\
 M_R M_I^\dagger &=& i \left( \begin{array}{ccc} -B_1 C_1 -B_2 C_2  & A_1 C_1- B_2 C_3  &  A_1 C_2+B_1 C_3  \\ -A_2 C_1-B_3 C_2  & B_1 C_1- B_3 C_3  &  B_1 C_2+A_2 C_3  \\ -B_3 C_1-A_3 C_2  & ~ B_2 C_1-A_3 C_3  &  B_2 C_2+B_3 C_3 \end{array}\right).
\end{eqnarray}

The diagonal elements give us following conditions
\begin{equation}
B_1 C_1 =-B_2 C_2 =B_3 C_3
\end{equation}
and the off-diagonal ones give us other three
\begin{eqnarray}
(A_1-A_2) &=& ~~~(B_3 C_2+B_2 C_3)/ C_1,  \\
(A_3-A_1) &=& ~~~(B_1 C_3-B_3 C_1)/ C_2,  \\
(A_2 -A_3) &=& -(B_2 C_1+B_1 C_2)/ C_3.
\end{eqnarray}
But, substituting Equation (11) into the sum of Equation (12) and Equation (13) will receive Equation (14).
So we have in fact only four equations to reduce the number of independent parameters down to  five.  \\

For simplicity, one may leave $A_3$, $B_3$, $C_3$, $B_1$ and $B_2$ independent and replaces $A_1$, $A_2$, $C_1$ and $C_2$ by
\begin{eqnarray}
A_1 &=& A_3 +B_2 (B_1^2 -B_3^2)/ B_1 B_3,\nonumber \\
A_2 &=& A_3 +B_3 (B_1^2 -B_2^2)/ B_1 B_2, \nonumber \\
C_1 &=& B_3 C_3 /B_1,~~~~~~
C_2 = -B_3 C_3 /B_2 ,
\end{eqnarray}
and Equation (5) now becomes
\begin{eqnarray}
M = \left( \begin{array}{ccc}
A + x B (y- {1 \over y})     & y B +i {C \over y}     & x B-i {C \over x}   \\
y B -i {C \over y}                 & A +B ({y \over x}-{x \over y})  &  B+i C \\
x B + i {C \over x}   & B -i C               & A   \end{array}\right),
\end{eqnarray}
if one lets $A \equiv A_3$, $B \equiv B_3$, $C \equiv C_3$ and $x \equiv B_2 / B_3$, $y \equiv B_1 / B_3$. \\

Subsequently, the mass eigenvalues can be derived analytically as
  \begin{eqnarray}
M^{diag.} =
 \left( \begin{array}{ccc}  A-B {x \over y} -C {\sqrt{x^2 +y^2 +x^2 y^2} \over {x y}} ~~~~~~~~~~ 0 ~~~~~~~~~~~ 0 \\
  0 ~~~~~~~~~ A-B{x \over y} + C{\sqrt{x^2 +y^2 +x^2 y^2} \over {x y}} ~~~~~~~~~~~ 0 \\
  0 ~~~~~~~~~~~~~~~~~~ 0 ~~~~~~~~~~~~~~~~~~~~ A+B{{(x^2+1) y} \over x} \end{array}\right),
\end{eqnarray}
with the $U$ matrix given as
\begin{eqnarray}
& ~~U^{(u)} & \\ \nonumber
&=& \left( \begin{array}{ccc}
{-\sqrt{x^2+y^2} \over \sqrt{2(x^2+y^2+x^2 y^2)}} &~ {{x(y^2-i \sqrt{x^2+y^2+x^2 y^2})} \over {\sqrt{2} \sqrt{x^2+y^2} \sqrt{x^2+y^2+x^2 y^2}}} &~{{y(x^2+i \sqrt{x^2+y^2+x^2 y^2})} \over {\sqrt{2} \sqrt{x^2+y^2} \sqrt{x^2+y^2+x^2 y^2}}}  \\
 {-\sqrt{x^2+y^2} \over \sqrt{2(x^2+y^2+x^2 y^2)}} &~ {{x(y^2+i \sqrt{x^2+y^2+x^2 y^2})} \over {\sqrt{2} \sqrt{x^2+y^2} \sqrt{x^2+y^2+x^2 y^2}}} &~{{y(x^2-i \sqrt{x^2+y^2+x^2 y^2})} \over {\sqrt{2} \sqrt{x^2+y^2} \sqrt{x^2+y^2+x^2 y^2}}} \\
 {{x y} \over \sqrt{x^2+y^2+x^2 y^2}} &~ {y \over \sqrt{x^2+y^2+x^2 y^2}} &~{x \over \sqrt{x^2+y^2+x^2 y^2}} \end{array}\right),
\end{eqnarray}
if we consider here the up-type quarks. \\

Surprisingly, all elements of this matrix are independent of A, B and C.
They depend only on two parameters $x$ and $y$.
Similarly, the matrix $U^{(d)}$ for down-type quarks should has the same pattern and one may express it simply by replacing all parameters in $M^{(u)}$ and $U^{(u)}$ with primed ones $A'$, $B'$, $C'$, $x' \equiv B'_2 / B'_3$ and $y' \equiv B'_1 / B'_3$, respectively. \\

Though this pattern was achieved by dividing $M$ into two real and imaginary components.
However, any combinations of $M_R$ and $M_I$ like $(p M_R +q M_I)$ with $p$ and $q$ arbitrary numbers, even complex, can be assigned to $M_1$ or $M_2$ and still be diagonalized by the same $U$.
Thus, any matrix pairs
\begin{eqnarray}
M_1=p M_R +q M_I ~~~{\rm and}~~M_2 = r M_R +s M_I ,
\end{eqnarray}
where $p,~q,~r$ and $s$ are arbitrary numbers, will also be diagonalized by the same $U$ and free of tree-level FCNCs naturally.
Besides, the derivation demonstrated above did not employ any symmetries like $Z_2$ in ~\cite{Glashow1977}, $S_3$ in ~\cite{Lin1988, Lee1990, Lin1994} or $(S_3 +S_2)$ in ~\cite{Lin2013} except for the only assumption of hermiticity of quark mass matrices.
Such a matrix pattern is a very general one.
Even, it includes all previous NFC models as special cases in it. \\

For instance, in Type-I and -II models some of the $M^{q}_{i}$ components are assigned zero by the $Z_2$ discrete symmetry.
That can be achieved by letting either $A=B=0$ or $C=0$ for corresponding quark type.
The $S_3$ pattern in Equation (1) can be achieved by letting $x=y=1$ (firstly achieved in ~\cite{Lin1988, Lee1990, Lin1994} or the case-1 in ~\cite{Lin2013}).
Those three patterns achieved in the $(S_3+S_2)$ model correspond to $x=y=-1$ (case-2), $x=-y=1$ (case-3) and $x=-y=-1$ (case-4), respectively.
Even the Fritzsch ansatz ~\cite{Fritzsch1978, Fritzsch1979} and the aligned two-Higgs doublet model (ATHDM) are also included.
One may achieve the Fritzsch ansatz simply by letting $A_1 =A_2= B_2 =0$.
For ATHDM, that is the $p \cdot s = q\cdot r$ case in this model.  \\

\section{Conclusions and Discussions}

The matrix pattern achieved in this article is a very general one since no symmetry is employed during the derivation.
The only assumption employed is the hermiticity of quark mass matrices.
Besides, it is derived analytically from an very fundamental theory of electro-weak interactions.
Thus, it includes almost previous 2HDMs in it, except for the type-III models.
That provides us with a very rational and general aspect to realize the nature of FCNCs and CP violation. \\

In this model all possible freedom in the Yukawa sector gets parameterized in terms of five parameters $A$, $B$, $C$ $x$ and $y$ for a quark type.
But, corresponding unitary transformation matrix $U^{(u)}$ depends only on two of them.
Assuming $U^{(d)}$ has similar FCNC-free pattern and assigning its corresponding parameters as $x'$ and $y'$, the CKM matrix thus derived will depend on merely three of the four $x$, $y$, $x'$ and $y'$ parameters if one considers the unitarity of CKM matrix.
That provides us with a new aspect to realize the explicit origin of CP violation. \\

Theoretically, of course, there could be other FCNC-free matrix pairs than those presented in this article.
For instance, there could be matrix pairs which are not respectively Hermitian but still can be diagonalized simultaneously.
Surely they will not satisfy Equation (7) and the number of free parameters in them will be many more than five for each fermion type. \\

Besides, the mass matrix $M$ itself could be non-Hermitian, too.
In that case, the number of free parameters surely will be many more than five.
Both cases are far beyond our ability of analytical derivation for now. \\

Beyond the 2HDM scope of this article.
It is interesting that these matrix patterns also apply to Standard Model and an even extended model with three Higgs doublets.
The detailed study on them is now still underway. \\

\end{document}